\documentclass{elsart}

\usepackage{graphicx}
\usepackage{dcolumn}
\usepackage{amsmath}

\usepackage{amssymb}
\begin{document}
\begin{frontmatter}

To be published in Carbon (2004) [doi:10.1016/j.carbon.2004.07.020]\\

\title{Searching for a magnetic proximity effect in magnetite-carbon structures}
\author{R.\ H{\"o}hne,}
\author{M.\ Ziese, } \author{P.\ Esquinazi}
\ead{esquin@physik.uni-leipzig.de} \address{Division of
Superconductivity and Magnetism,University of Leipzig,
Linn\'estrasse 5,  04103 Leipzig, Germany}


\begin{abstract}
In order to study a possible magnetic proximity effect in magnetite-carbon
structures, we have performed magnetization measurements of
graphite-magnetite composites with different mass ratios as well as the
measurement of the magnetoresistance of one of them and of the
magnetization of a magnetite-carbon bilayer. The overall results do not
indicate the induction of bulk ferromagnetism in graphite and disordered
carbon structures through their contact with magnetite.
\end{abstract}
\end{frontmatter}
\section{Introduction}
The possibility of magnetic ordering in metal-free carbon structures with
Curie temperatures above room temperature attracts the interest of a broad
spectrum of natural sciences. First systematic indications were obtained
more than ten years ago by Japanese groups but without triggering the
interest of the scientific community \cite{murata91,murata92}. The
ferromagnetism observed in C$_{60}$ polymerized in oxygen atmosphere by
light\cite{murakami96} and by annealing at high temperatures and pressures
\cite{makanat01,hancar03,hanadd}, as well as in proton irradiated highly
oriented pyrolytic graphite \cite{pabloprl03} provides indications for a
metal-free magnetic ordering above room temperature.

Nevertheless, the influence of ferromagnetic impurities, especially that
of Fe and magnetite (Fe$_3$O$_4$), has to be taken into account carefully
because most of the carbon-based samples show relatively weak
ferromagnetic signals. Measurements of the magnetization of graphite
nodules from a meteorite\cite{coey02} and of the magnetic force gradients
of carbon nanotubes in contact with ferromagnetic
substrates\cite{cespedes04} suggest that magnetic ordering in graphite and
probably in other carbon-based structures\cite{makareview} might be
induced by an anomalously large magnetic proximity effect. A relatively
large magnetization-decay length (identical to the spin-diffusion length)
$\lambda_s \approx 5~$nm was estimated for the interface
graphite-magnetite \cite{coey02}.  Without doubt, a magnetic proximity
effect in carbon structures with such a decay length would provide
interesting possibilities for applications as well as trigger further
theoretical and experimental work. Therefore, further systematic studies
are necessary to check whether such a phenomenon exists and how large
$\lambda_s$ might be. The present work is a  systematic approach to
estimate experimentally the possible magnetization-decay length. For that
purpose we have measured the magnetization of graphite-magnetite
composites with different ratios of the two components and the
magnetoresistance of one of them.

We note that the graphite matrix in the meteorite nodules measured
in Ref.~\cite{coey02} is unlikely to be well ordered. The
magnetoresistance of the meteorite graphite reported
  in Ref.~\cite{coey02} deviates qualitatively and quantitatively
  from that measured in highly oriented samples \cite{yakovadv03}.
  The observed square field dependence as well as the
   very small increase of the resistance with a field of a few tesla\cite{coey02}
   indicate that the
  graphite material in the meteorite must be considered as highly disordered.
  Taking this into account we
have tried to simulate an interface between magnetite and a highly
disordered graphite material by depositing  a disordered carbon
film on top of a magnetite film. Within experimental errors, the
results obtained for the magnetization of our samples as well as
magnetoresistance measurements performed on the graphite-magnetite
composites do not provide clear indication that such a large
proximity effect exists. Taking into account these results we
suggest other possibilities for the origin of the additional
magnetization measured in the graphite nodules.

The paper is organized as follows. In the next section details on sample
preparation and experimental methods are given. In section \ref{results}
we present the results with the discussion and in section \ref{conclu} the
conclusion with an alternative explanation for the observed ferromagnetic
response of the meteorite nodule.

\section{Experimental details \label{experimental}}
A magnetite film was prepared by pulsed laser deposition (PLD)
from a stoichiometric polycrystalline magnetite target onto a
(100) MgAl$_2$O$_4$ substrate with an area of $5 \times 5$~mm$^2$.
Substrate temperature and oxygen partial pressure during
deposition were $430^\circ$C and $1\times 10^{-5}$~mbar,
respectively. The laser pulse energy was 600~mJ; the film was made
with 6800 laser pulses corresponding to a magnetite film thickness
of $\simeq 20$~nm. After characterization of the magnetite film a
carbon film was deposited onto this film by PLD in another vacuum
chamber especially used for nonmagnetic materials. The deposition
was performed at room temperature, in vacuum ($2\times
10^{-5}$~mbar) with a laser pulse energy of 800~mJ and a
repetition rate of 3~Hz.
 4000 pulses were used to get a thickness of the carbon film of
$(15\pm 5)$~nm.

The composites  were fabricated from commercial Fe$_3$O$_4$ and
graphite powder. Both graphite (magnetic impurity content $<
0.1$~ppm) and magnetite powder had micron sized grains. The
powders were mixed in an appropriate mass ratio in an agate mortar
and afterwards pressed into pellets with radius of 5~mm and
typical height of 2~mm with a typical mass of $\sim 30$~mg.

The magnetic moment of the samples was measured with a  SQUID magnetometer
(MPMS-7, Quantum Design). In case of the composites the measurements were
carried out in standard DC mode. For the bilayer the measurements were
performed before and after the carbon deposition using the reciprocating
sample option (RSO) with a sensitivity of $\sim 10^{-10}$~Am$^2$.
Resistance measurements were performed in the van der Pauw configuration
\cite{vanderpauw1958} using a He-flow cryostat (Oxford Instruments)
equipped with a 9~T superconducting solenoid. The current-voltage
characteristics were linear at least up to a current of 100~mA.
\section{Experimental Results and Discussion \label{results}}
\subsection{Magnetite-Carbon Bilayer}\label{bilayer}

\begin{figure}
\begin{center}
\includegraphics[width=0.8\textwidth]{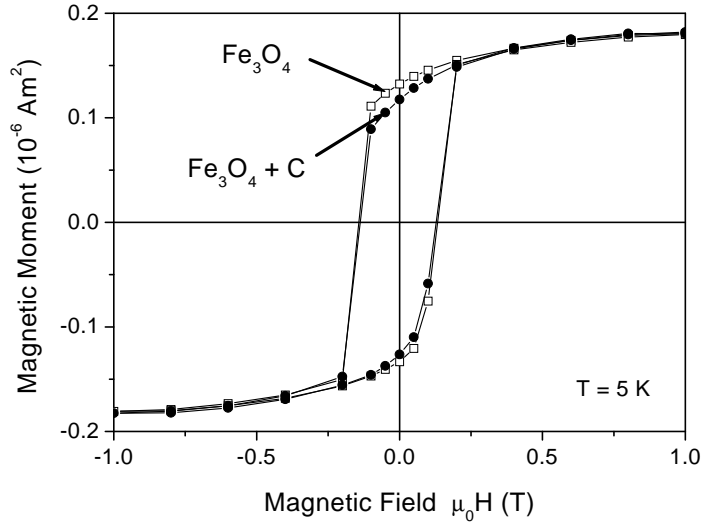}
\end{center}
\vspace{0.0cm} \caption{The magnetic moment as a function of
applied field parallel to the main area of the bilayer
Fe$_3$O$_4$-C. Within 1\% the values of the magnetic moment at
$\mu_0 H = 1~$T of magnetite and of the bilayer are similar. It is
noticeable that the values at lower fields $\mu_0H < 0.2$~T are
smaller for the bilayer, in the case of the remanence 5\% in the
negative and 12\% in the positive part, as for the magnetite film
without the carbon layer.} \label{h1}
\end{figure}

\begin{figure}
\begin{center}
\includegraphics[width=0.8\textwidth]{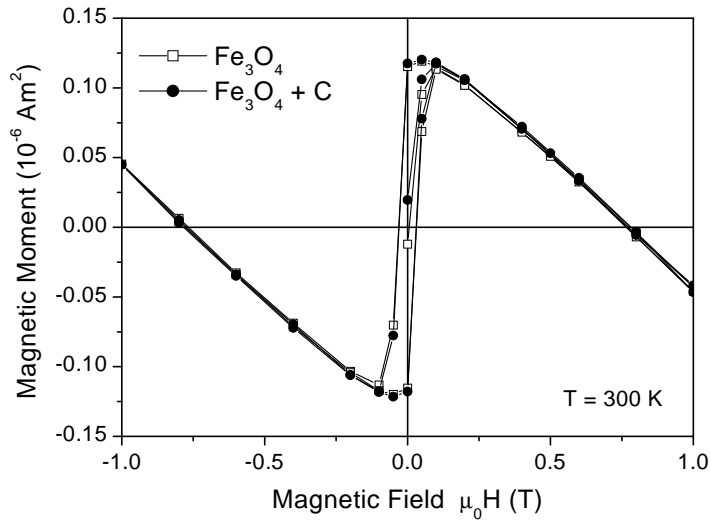}
\end{center}
\vspace{0.0cm}
\caption{Fig.~\protect{\ref{h2}} shows the same as Fig.~\protect{\ref{h1}}, but for
$T = 300$~K.}
\label{h2}
\end{figure}

\begin{figure}
\begin{center}
\includegraphics[width=0.8\textwidth]{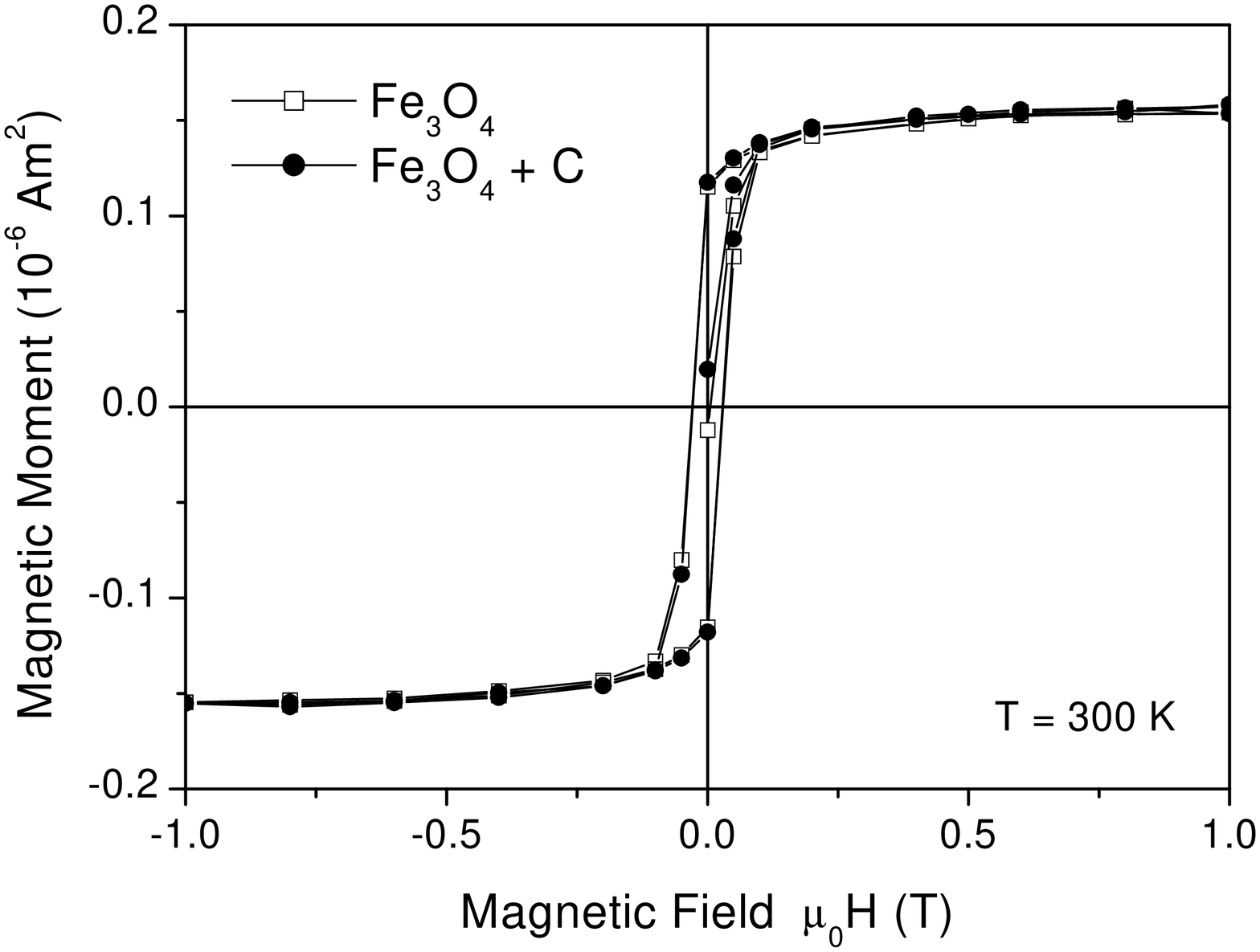}
\end{center}
\vspace{0.0cm} \caption{Fig.~\protect{\ref{h3}} shows the same as
Fig.~\protect{\ref{h2}}, but after subtraction of the diamagnetic
contribution from the substrate.} \label{h3}
\end{figure}

Figure~\ref{h1} shows the magnetic moment of the magnetite film alone and
of the same sample after carbon deposition as a function of magnetic field
by cycling the field between $+1$~T and $-1$~T at a temperature $T = 5$~K.
Figures~\ref{h2} and \ref{h3} show similar measurements but at $T =
300~$K.
The values of the saturation magnetic moment measured at $T =
300$~K are similar within 0.4\%  for magnetite alone and the
bilayer.  At this temperature the remanence value is about 2\%
larger in the bilayer than in the magnetite film alone. At 5~K
however, the remanence decreased after depositing the carbon film,
see Fig.~\ref{h1}. Our experience indicates that the
reproducibility of the measurements in our SQUID is within 0.5\%.
Therefore, the small differences in the magnetic saturation
moments between Fe$_3$O$_4$ and the Fe$_3$O$_4$/C bilayer are near
the experimental error limit and no clear magnetic influence can
be asserted to Fe$_3$O$_4$ on carbon.

Nevertheless, a maximum estimate for the penetration depth can be
obtained for the bilayer system within experimental error. The
influence of the carbon layer on the magnetization of the bilayer
without a proximity effect can be estimated as follows. The mass
of the carbon layer for $\simeq 15$~nm thickness and a mass
density of 2~g/cm$^3$ is $\simeq 10^{-6}$~g. The paramagnetic
contribution of the carbon film\cite{hohcfilms} to the magnetic
moment of the Fe$_3$O$_4$/C bilayer at $\mu_0H = 1$~T and $T =
5$~K is of the order $1\dots 2 \times 10^{-7}$~emu ($1\dots 2
\times 10^{-10}$Am$^2$), i.e.~three orders of magnitude smaller
than the corresponding value of the sample. At room temperature
the contribution of the carbon layer is even smaller.

Let us assume now naively an effective magnetic penetration depth of
magnetite in the carbon layer of 0.4~nm (about half of the lattice
constant of magnetite) and that in this depth the carbon material shows
the same magnetization as magnetite. In this case 2\% of the magnetite
volume is composed of carbon and the saturation magnetic moment of the
bilayer would be enhanced by 2\% in comparison to the single magnetite
film. Within experimental error the observed maximum possible increase of
the magnetization of the bilayer is 1\% and 0.4\% at 5~K and 300~K,
respectively. In this case the observed changes in the saturation moments
would correspond to effective penetration depths  of 0.2~nm and 0.08~nm,
respectively.

Let us assume an exponentially decaying magnetization in the carbon film
of the form $M = M_0 \exp(-z/\lambda_s)$. In this case the magnetic moment
in the carbon film of area $A$ and thickness $d_c$ is given by
\begin{equation}
m_{\rm carbon} = M_0 A \int_0^{d_c}\, dz\, \exp[-z/\lambda_s] =
M_0 A \lambda_s\left[1-\exp(-d_c/\lambda_s)\right] \label{moment3}
\end{equation}
and the ratio of magnetic moments of the carbon and the magnetite
film (thickness $d_F$) is
\begin{equation}
\frac{m_{\rm carbon}}{m_{\rm magnetite}} = \frac{\lambda_s}{d_F}\,
\left[1-\exp\left(-\, \frac{d_c}{\lambda_s}\right)\right]\, .
\label{moment4}
\end{equation}
With $d_c = 15$~nm and $d_F = 20$~nm and from the remanence values at
300~K we may write $m_{\rm carbon}/m_{\rm magnetite} \le 0.02$. From
Eq.~(\ref{moment4}) we obtain an upper limit for the magnetic-diffusion
length of magnetite in carbon $\lambda_s \precsim 0.4$~nm. If the
penetration depth of magnetite in carbon  would be 5~nm, as assumed in
Ref.~\cite{coey02} the enhancement would be easily measurable. This is not
observed experimentally.

\subsection{Magnetization of Magnetite-Graphite Composites}

\begin{figure}
\begin{center}
\includegraphics[width=0.8\textwidth]{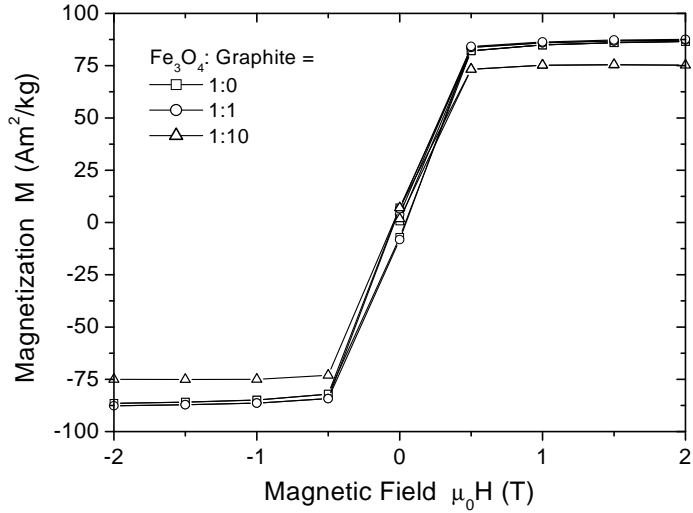}
\end{center}
\vspace{0.0cm} \caption{Magnetization of a pure magnetite powder compact
and magnetite-graphite composites with mass ratios 1:1 and 1:10. The
magnetization was calculated from the measured magnetic moment by dividing
through the magnetite mass alone. The measurements were carried out at
130~K.} \label{z1}
\end{figure}

The magnetization curves of a pure magnetite powder compact and
magnetite-graphite composites with mass ratios 1:1 and 1:10 are shown in
Fig.~\ref{z1}. The measurements were performed at 130~K in order to work
at low temperatures but above the Verwey transition of magnetite. The mass
magnetization was calculated from the measured magnetic moment divided by
the {\em magnetite weight alone}. In case of the 1:1 composite the
magnetization increases $\le 1.2\%$ at 1~T, whereas the magnetization of
the 1:10 composite clearly decreases. Only a part of this decrease could
be due to the inaccurate magnetite mass determination in this dilute
mixture. Note that quantitatively  the observed changes cannot be due to
the diamagnetic contribution of the graphite powder because this shows a
magnetization at 130~K of\cite{hohcfilms} $M_{\rm graphite} =
-0.01$Am$^2$/kg
 and it contributes, even  for the 1:10 composite, by a
factor less than $1.5 \times 10^{-3}$ to the total magnetic moment. Taking
into account a relative accuracy of the order of 0.5\% between different
runs in the SQUID (measuring the same sample re-mounted in another sample
holder) and a mass accuracy of 1\%, the magnetization increase observed
for the 1:1 composite is within experimental error. Due to the relatively
large size of the magnetite grains, this experimental error is already too
large to definitely rule out the existence of an anomalous large proximity
effect as the estimates done below show.

We can estimate the maximum spin-diffusion length as follows. Consider a
spherical magnetite particle with radius $r_0$ and magnetization $M_0$
embedded in graphite. This is assumed to induce a magnetization in the
adjacent graphite of the form $M_0\exp[-(r-r_0)/\lambda_s]$. Here the
strong anisotropy of graphite is neglected such that $\lambda_s$ is some
orientation-averaged quantity. The magnetic moment in the graphite can be
calculated to
\begin{equation}
m_{\rm graphite} = 4\pi M_0\, \int_{r_0}^\infty\, dr\,
r^2\exp[-(r-r_0)/\lambda_s] = 4\pi
M_0\left[r_0^2\lambda_s+2r_0\lambda_s^2+2\lambda_s^3\right]
\label{moment1}
\end{equation}
and the ratio of the magnetization of the graphite and the magnetite
inclusion is then given by (for a mass ratio 1:1)
\begin{equation}
\frac{M_{\rm graphite}}{M_0} = 3u\left[1+2u+2u^2\right] \label{moment2}
\end{equation}
with $u = \lambda_s/r_0$. If we assume that for the composite with
mass ratio 1:1 a magnetization enhancement of 1.2\%, $u \lesssim
0.004$. With a typical magnetite-particle radius of $1$~$\mu$m
this sets an upper limit to the spin-diffusion length of
$\lambda_s \lesssim~$4~nm. This value is within the value assumed
in Ref.~\cite{coey02} although the experimental error does not
allow a clearer statement. In the case of the 1:10 composite, the
total absence of an increase in the magnetization suggests a
negligible  spin penetration depth of magnetite in the graphite
grains.

\subsection{Magnetoresistance of Magnetite-Graphite Composites}

\begin{figure}
\begin{center}
\includegraphics[width=0.8\textwidth]{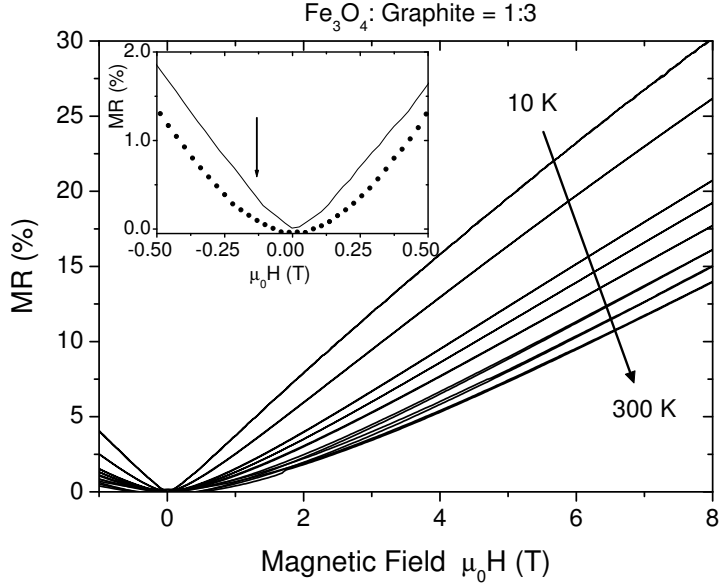}
\end{center}
\vspace{0.0cm} \caption{Magnetoresistance ratio of a
magnetite-graphite composite with mass ratio 1:3 at various
temperatures, from top to bottom: 10, 50, 100, 120, 150, 200, 250,
300~K. The inset shows the magnetoresistance at 10~K (continuous
line) and 200~K (points) on an expanded scale. The arrow indicates
the position of the coercive field of magnetite powder at 10~K.}
\label{z2}
\end{figure}

If some magnetization is induced in the graphite grains, one might
expect to observe some anomalies near the coercive field in the
magnetoresistance of the composite. Therefore as an additional
evidence we performed magnetoresistance measurements on a
magnetite-graphite composite with mass ratio 1:3. The
magnetoresistance ratio is defined as
\begin{equation}
MR = \frac{R(H)-R(0)}{R(0)}
\label{mr}
\end{equation}
and it is shown in Fig.~\ref{z2} for this composite. With increasing
temperature the magnetoresistance decreases and the functional form
changes from linear to quadratic as it is typical for graphite. An effect
of the magnetite grains on the total magnetoresistance could not be
detected. This becomes more evident from the inset which shows the
magnetoresistance at two temperatures on an expanded field scale.
\section{Conclusion and alternative explanations \label{conclu}}
The results presented in this experimental study do not indicate a
significant enhancement of the ferromagnetism induced by magnetite in bulk
carbon. Although in the case of the composites of mass ratio 1:1 and
because of the relatively large graphite grains embedded in a magnetite
matrix (or vice versa)  the experimental error does not allow to rule out
completely a spin-diffusion length of the order of a few nanometers, the
overall results do not seem to support this picture. It is clear that more
experiments have to be done, especially using large and thin
ordered-graphite - magnetite interfaces for a complete and accurate
characterization of an anomalously large proximity effect. This task is
however not simple due to the difficulties one has to grow one material
onto the other. Recently, local ferromagnetism in C/Fe multilayers was
observed by resonant magnetic reflectivity of circularly polarized
synchrotron radiation \cite{mertins04}. The thickness of the carbon layers
was 0.55~nm in comparison with 2.55~nm for the Fe-layers. Taking into
account an average roughness of 0.35~nm at the interface where a mixture
of C and Fe exists, the induced ferromagnetism appears within a
penetration depth of less than 0.5~nm. This result would agree with the
results of Sec.~\ref{bilayer} where we obtain $\lambda_s \precsim 0.4~$nm.

The experimental data reported in Ref.~\cite{coey02} indicate that
at least two samples from the graphite nodule do not show an
enhancement of ferromagnetism in the carbon matrix, although more
than 40\% (in weight) graphite was in contact with almost the same
mass quantity of Fe and magnetite. There is no clear reason why
the proximity effect between graphite and the Fe-based
ferromagnets should not apply for all the samples reported in
Ref.~\cite{coey02}. On the other hand, those results show an
inhomogeneously distributed  magnetic ordering in the carbon
matrix of the meteorite. Taking into account these and our results
we suggest an alternative explanation for the ferromagnetism
observed in the meteorite.

Previous\cite{murata92} and recently published experimental
work\cite{pabloprl03} indicate that hydrogen may trigger a ferromagnetic
ordering in graphite. Theoretical work\cite{kusakabe03,maru04} suggests
that a mixture of sp$^2$ and sp$^3$ carbon-hydrogen bonds may shift the
spin-band structure of a graphene layer producing a 100\% spin polarized
band. Other theoretical work\cite{lehtinen04} indicates the existence of
an extraordinarily large magnetic moment localized at a carbon vacancy
position when a H atom occupies this place. Recently published work
indicates that hydrogen modifies substantially the electronic structure of
graphite around it. According to STM/AFM measurements \cite{rufi00} a
single H-atom interacting with a graphite surface modifies the electronic
structure over a distance of 20 to 25 lattice constants. Also, muon spin
rotation/relaxation experiments \cite{cha02} indicate that a positive muon
(basically equivalent to a proton) in graphite triggers a local magnetic
moment around it. We note further that it is highly probable that the
interstellar matter studied in Ref.~\cite{coey02} has been naturally
proton irradiated in space. Astrophysicists have speculated already some
time ago that interstellar graphite and other carbon-based meteorites and
dust may act as catalyst for the creation of molecular hydrogen in the
universe (see Ref.~\cite{jeloaica99} and Refs. therein). Therefore, the
measurement of the H content of those samples and the behavior of the
ferromagnetic signal as a function of high-temperature annealing time
could bring light on the origin of their ferromagnetism. As a further
example we refer to the ferromagnetism in polymerized fullerenes, where an
inhomogeneously distributed ferromagnetic material (only $\sim 30\%$ of a
ferromagnetic sample appears to be ferromagnetic), is observed by magnetic
force microscopy measurements \cite{hancar03,hanadd}. A defect structure
added to the influence of hydrogen or other light atoms, may lead to the
observed inhomogeneous magnetic ordering.

If the ferromagnetism in the graphite matrix is related neither to
disorder nor to the bonding of H (or other light atom), then it may be
possible that embedded Fe in the graphite matrix triggers an unusual
magnetic ordering in the graphite band structure. Recent calculations of
the magnetic moment of Fe atoms fixed at specific places of the graphene
layer indicate a slight decrease of its value compare to that of the free
atom \cite{yagi04}. However, to our knowledge it appears that there is no
calculations on the influence of Fe atoms on the spin distribution of the
graphene layer if these reside  at specific sites, e.g. at the edges or at
the vacancies of the structure. Theoretical studies in this direction
would be very helpful to understand the influence of magnetic ions in
carbon structures. From the experimental side irradiation of graphite
samples with Fe ions may provide some light on this issue. These
experiments are currently being performed and their results will be
published elsewhere.

\noindent {\bf Acknowledgements}

This work was supported by the DFG under Contract No.\ DFG ES 86/7-3
within the Forschergruppe ``Oxidi\-sche Grenzfl\"achen'' and under
Contract No.\ DFG ES 86/6-3.

%

\end{document}